\documentclass[cameraready]{Interspeech}
\usepackage{amsmath,graphicx,hyperref}
\usepackage{xcolor,soul,framed} 
\colorlet{shadecolor}{yellow}
\usepackage{array}
\usepackage{url}
\usepackage{cite}
\usepackage{booktabs}
\usepackage{balance} 
\usepackage{multirow}
\usepackage{multicol}
\usepackage{amsfonts}
\usepackage{float}
\usepackage{bbold}
\usepackage{hyperref}
\usepackage{etoolbox,siunitx}
\robustify\bfseries
\usepackage{balance}
\usepackage{amssymb}
\usepackage{pifont}
\usepackage{scalerel}

\title{TF-MossFormer: Integrating Convolution Gated Local-Global Attentions for Enhanced Time-Frequency Domain Monaural Speech Separation}

\author{Shengkui}{Zhao}
\author{Zexu}{Pan}
\author{Haoxu}{Wang}
\author{Biao}{Tian}
\author{Bin}{Ma}
\author{Xiangang}{Li}

\address{
    Token Foundry, Alibaba Group, Singapore 
}

\email{shengkui.zhao@alibaba-inc.com}


\usepackage{comment}

\begin{document}

\maketitle

\begin{abstract}
Transformers with global attention capture long-range dependencies but can miss the fine-grained local continuity crucial for speech separation. We propose TF-MossFormer, a time–frequency transformer that combines local and global attention to jointly model short- and long-range contexts for monaural speech separation. At its core is a content-aware sliding-window attention mechanism that dynamically adapts receptive fields for stronger local interactions, avoiding the rigidity of static convolutions. Unlike time-domain chunk-based methods, TF-MossFormer leverages the 2D spectrogram to model structure along both time and frequency axes. Convolutional gating between attention layers further improves feature selection and information flow. TF-MossFormer achieves SI-SDRi of 22.6, 24.0, and 24.4 dB on WSJ0-2Mix with 5.9M, 16.9M, and 25.4M parameters, respectively, outperforming prior approaches.
\end{abstract}

\section{Introduction}
\label{sec:intro}

Speech separation (SS), which aims to isolate individual sources from a mixture, is a fundamental challenge in audio processing. Recent research reveals that effective modeling requires capturing both short-range continuity, such as harmonic structures and phoneme-level transitions, and long-range grouping, including speaker identity consistency across an utterance \cite{Erdogan2015, Luo2018N, Luo2018Real, Chen2020Q}. Early end-to-end approaches like Conv-TasNet \cite{Luo2019} highlighted the importance of local feature extraction through convolutional networks. This idea was extended by the time-domain dual-path architecture DPRNN \cite{Luo2020Z}, which explicitly decomposed the task into intra-chunk (local) and inter-chunk (global) modeling to jointly address short- and long-range dependencies. Transformer-based models such as SepFormer \cite{Subakan2021M} and our previous MossFormer \cite{Zhao2023M} and MossFormer2 \cite{Zhao2024M} further advanced this line of work by leveraging self-attention to enhance global context modeling, leading to remarkably improved separation performance.
\begin{figure}[t]
  \centering
  \includegraphics[width=7.6cm]{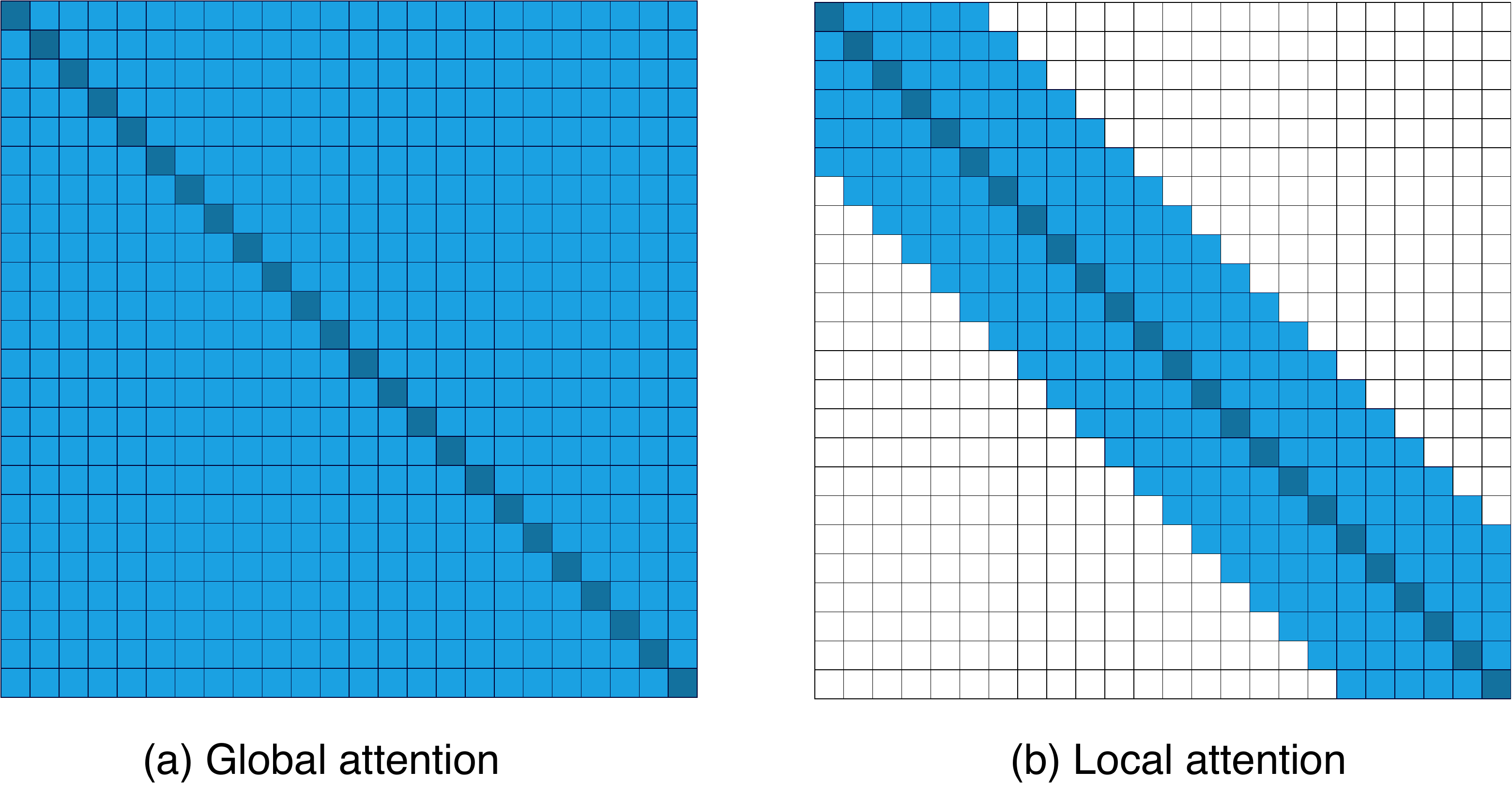}
  \vspace{-3mm}
  \caption{Illustration of the global and local attention patterns in TF-MossFormer. (a) Global attention: captures long-range dependencies across the entire spectrogram, (b) Local attention: focuses on fine-grained continuity within sliding windows.}
  \label{fig1}
\vspace{-3mm}
\end{figure}

Compared with time‑domain dual‑path methods, time–frequency (TF) approaches offer a distinct advantage by leveraging the spectrogram’s two‑dimensional structure \cite{chen2021continue, Yang2022L, wang2023tfgridnet, li2024spmamba, Kohei2024tflocoformer}. For example, TF‑GridNet \cite{wang2023tfgridnet} outperformed previous time‑domain models by explicitly encoding spectrograms into separate temporal and frequency processing paths, using bidirectional LSTMs to preserve fine‑grained continuity and full‑band self‑attention to capture long‑range dependencies across the entire spectrogram. Building on this framework, SPMamba \cite{li2024spmamba} replaced recurrent blocks with state‑space models, showing that alternative sequential operators can deliver better local modeling at lower cost. Conformer‑based methods \cite{chen2021continue} take a complementary approach by embedding convolutional modules inside attention blocks, explicitly fusing locality and context to improve the modeling of short‑range transients. TF‑LocoFormer \cite{Kohei2024tflocoformer} further complements global attention with gated convolutional layers to more tightly integrate local spectral detail and global aggregation, yielding notable gains in separation performance. These prior works motivate our focus on adaptive local mechanisms that enhance spectral continuity while preserving global modeling capabilities based on the TF domain geometry.

The shift toward hybrid local–global modeling is a broader trend beyond speech processing. In natural language processing (NLP), models like Longformer \cite{Beltagy2020Longformer} and BigBird \cite{zaheer2020bigbird} combine local attention with global tokens to efficiently handle long sequences. Similarly, in computer vision, PLG-ViT \cite{ebert2023light} and BiFormer \cite{zhu2023biformer} use parallel attention to capture both fine details and scene-level context. These approaches reflect a broader need to model both local patterns and long-range dependencies in sequential data.
\begin{figure*}
  \centering
  \includegraphics[width=15.0cm]{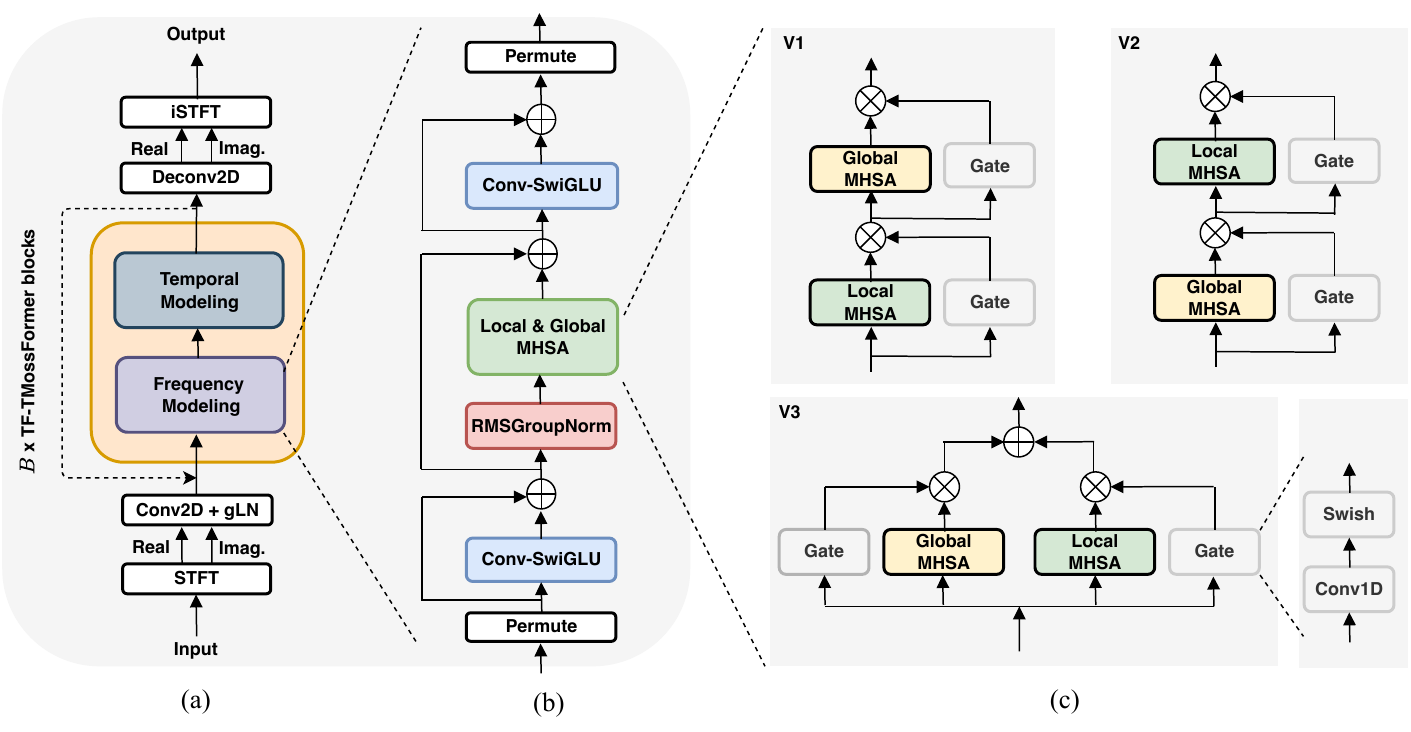}
  \vspace{-3mm}
  \caption{Overview of TF-MossFormer. (a) Model flowchart with STFT/ISTFT, Conv2D/Deconv2D, gLN, and $B$ stacked TF-MossFormer blocks for interleaved time–frequency modeling. (b) Configuration of the frequency modeling block (the temporal block adopts the same structure). (c) Three variants (V1–V3) of the gated local–global MHSA module. $\oplus$: element-wise addition, $\otimes$: element-wise multiplication.}
  \label{fig2}
\vspace{-3mm}
\end{figure*}
Inspired by these cross-domain innovations, we introduce TF-MossFormer, a time–frequency framework for monaural speech separation. Our model integrates hybrid local–global attention in a unified architecture to preserve short-range spectral continuity while capturing long-range dependencies in speech spectrograms. As shown in Fig. 1, sliding-window local attention focuses on fine-grained spectral features, retaining subtle temporal and harmonic details, while global attention spans wider regions to capture speaker-specific characteristics. We further enhance both pathways with convolutional gating to improve representation. We evaluate the performance of TF‑MossFormer through extensive experiments on the widely-used WSJ0‑2Mix benchmark.

\section{The TF-MossFormer Model}
\label{sec:format}
Given a speech mixture $x = \sum_{i=1}^{C} s_i$, our goal is to estimate the $C$ individual sources 
$s_i \in \mathbb{R}^{1 \times L}, \; i = 1, 2, \dots, C$ using a deep learning model. 
The proposed TF-MossFormer adopts the dual-path time--frequency separation paradigm~\cite{wang2023tfgridnet}, 
preserving the encoder--separator--decoder architecture. 
Unlike time-domain approaches that operate on two-dimensional chunks and predict time-domain masks, TF-MossFormer operates in the short-time Fourier transform (STFT) domain 
and directly predicts both the real and imaginary components of the complex spectra.
 
\subsection{Overview of TF-MossFormer Architecture}
Fig. \ref{fig2} illustrates our TF-MossFormer architecture. The input mixture waveform $x$ is first transformed into the time–frequency domain using STFT, producing real and imaginary components $\mathbf{X}_\text{Real} \in\mathbb{R}^{T\times F}$ and $\mathbf{X}_\text{Imag.} \in\mathbb{R}^{T\times F}$, where $T$ and $F$ are the number of frames and frequency bins. The real and imaginary parts are concatenated as $\mathbf{X} \in\mathbb{R}^{2\times T\times F}$ and passed through a 2D convolutional (Conv2D) encoder followed by global Layer Normalization (gLN) to project the features into a feature representative space.
The core separator consists of $B$ stacked TF-MossFormer blocks, each comprising two sub-modules: the frequency module and the temporal module. Both modules share the same architecture but perform on alternative sequence dimensions. The  frequency  module models spectral dependencies across frequency bins along the frequency path, while the temporal module models temporal dependencies across time frames along the time path.
By alternately modeling frequency and time dependencies, the network progressively refines the representation while efficiently covering the entire time–frequency context.
The decoder uses a transposed convolution (Deconv2D) layer to reconstruct the estimated real and imaginary spectra, which are then transformed back to the time domain using the inverse STFT (iSTFT) to obtain the separated waveform signals.

\subsection{The Encoder and Decoder}
Our TF-MossFormer adopts the encoder and decoder design from TF-GridNet. The encoder consists of a $3 \times 3$ 2D convolution (Conv2D) layer with $D$ output channels, followed by global layer normalization (gLN). This step projects each time–frequency (T–F) unit into a $D$-dimensional latent embedding, resulting in a tensor of shape $D \times T \times F$. 
The decoder mirrors the encoder by applying a $3 \times 3$ 2D transposed convolution (Deconv2D) with $2C$ output channels, representing the predicted real and imaginary components for all $C$ target sources. 

\subsection{The TF-MossFormer Separator}
The encoded representation is passed to the TF-MossFormer separator, which consists of $B$ stacked TF-MossFormer blocks. Within each block,  the spectro-temporal modeling is performed to progressively refine the local–global dependencies for the latent TF embeddings. As depicted in Fig. \ref{fig2}(a), each TF-MossFormer block contains two sequential modeling modules: a frequency modeling module and a temporal modeling module, both sharing an identical architecture. The module architecture is shown in Fig.~\ref{fig1}(b), where the input tensor is first permuted to align the modeling dimension (frequency or time) with the sequence axis. The permuted input is passed through a convolutional SwiGLU feed-forward network (Conv-SwiGLU) \cite{Kohei2024tflocoformer}, followed by a residual connection to preserve the original representation. Next, Root Mean Square Group Normalization (RMSGroupNorm) \cite{Kohei2024tflocoformer} is applied to stabilize training and improve convergence. The normalized features are then processed by the core Local \& Global Multi-Head Self-Attention (MHSA) module, which integrates content-aware local attention (via sliding windows) and global attention to jointly model fine-grained dependencies and long-range context. The output is again passed through a Conv-SwiGLU module and added back via residual connection before being permuted back to its original shape. This design ensures balanced local and global representation learning while maintaining efficiency through residual pathways and normalization.
\begin{figure}[t]
  \centering
  \includegraphics[width=6.6cm]{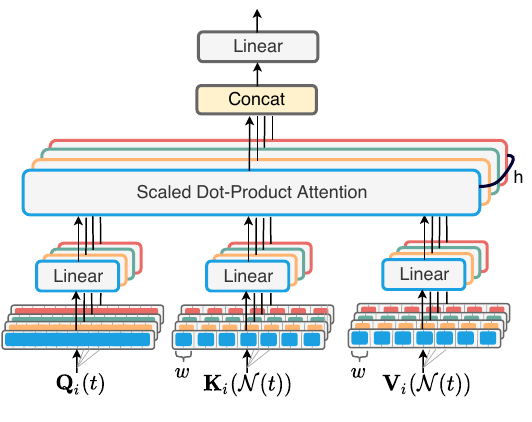}
  \vspace{-3mm}
  \caption{An overview of the local attention module, which performs multi-head self-attention using sliding windows. Each attention head operates on a fixed-size window of the input sequence.}
  \label{fig3}
\vspace{-3mm}
\end{figure}

\subsubsection{Convolution Gated Local and Global Attention}
We apply both local and global attention with convolution gating in the TF-MossFormer separator to capture complementary short- and long-range dependencies in speech spectrograms. Our global attention is a standard Multi-Head Self-Attention (MHSA) mechanism \cite{Vaswani2017N}, every frame in a sequence attends to every other frame as depicted in Fig. \ref{fig1}(a). For the temporal sequence $\mathbf{X}\in \mathbb{R}^{T\times D}$ T, the query ($\mathbf{Q}_i$), key ($\mathbf{K}_i$), and value ($\mathbf{V}_i$) matrices are the $i$-th head projections of size $T\times d_h$, where $d_h=D/h$ is the hidden dimension of each head. Our global MHSA computes the attention of each head as:
\begin{equation}
\mathrm{head}_{\text{global},i} = 
\mathrm{softmax}\left(\frac{Q_i K_i^{\top}}{\sqrt{d_h}}\right) V_i.
\end{equation}
All heads are concatenated to form the full attention matrix: 
\begin{equation}
\text{MHSA}_{\mathrm{global}}(\mathbf{X}) = 
\mathrm{Concat}\bigl(\mathrm{head}_{\text{global},1},\ldots,\mathrm{head}_{\text{global},h}\bigr)\mathbf{W}_O,
\end{equation}
where $\mathbf{W}_O$ is the output projection matrix. This operation has complexity $\mathcal{O}(T^2 D)$ due to the dense $QK^{\top}$ computation.

Complementing the global attention,  we present local attention to focus on short-range time–frequency patterns to preserve fine-grained continuity. As depicted in Fig. \ref{fig1}(b), it captures context at varying distances from each frame. Our local attention adopts the sliding-window multi-head attention as depicted in Fig. \ref{fig3}. The local attention restricts each frame $t$ to attend only to frames within a local window
$\mathcal{N}(t) = \{t-\frac{w-1}{2}, \ldots, t, \ldots, t+\frac{w-1}{2}\}$, where $w$ is the window size. 
The attention for head $i$ becomes:
\begin{equation}
\mathrm{head}_{\text{local},i}(t) = 
\mathrm{softmax}\left(
\frac{Q_i(t) K_i(\mathcal{N}(t))^{\top}}{\sqrt{d_h}}
\right)V_i(\mathcal{N}(t)),
\label{eqn:local1}
\end{equation}
where $Q_i(t) \in \mathbb{R}^{1 \times d_h}$ is the query for frame $t$, 
$K_i(\mathcal{N}(t)) \in \mathbb{R}^{w \times d_h}$ are the keys within the window, and $V_i(\mathcal{N}(t))$ are the corresponding values. 
This reduces complexity to $\mathcal{O}(T w D)$, which scales linearly with $T$ for a fixed $w$. 

We further augment local and global attention with convolutional gates as shown in Fig. 2(c). The gate consists of a Conv1D layer followed by Swish activation. Since attention ordering affects performance, we implemented three block variants (V1-V3) and ran an ablation study to identify the most effective arrangement for our model.
\begin{table}
\centering
\footnotesize
\caption{Hyperparameter notations and model configurations for TF-MambaFormer (S), (M), and (L).}
\setlength{\tabcolsep}{1.0pt}
\begin{tabular}{l lccc}
\toprule
Symbol    & Description & S & M &L \\
\midrule
$D$  & Embedding dimension of each TF bin & 96 & 128 & 128 \\
$B$  & Number of TF-MossFormer blocks    & 4 & 6 & 9 \\
$H$  & Hidden dimension in Conv-SwiGLU    & 256 & 384 & 384 \\
$O$  & Output dimension in Conv-Gate    & 96 & 128 & 128 \\
$K$  & Kernel size in Conv1D and Deconv1D & 4 & 4 & 4 \\
$S$  & Stride in Conv1D and Deconv1D      & 1 & 1 & 1 \\
$h$  & Number of heads in self-attention  & 4 & 4 & 4 \\
$w_F$  & Sliding-window length for frequency path  & 7 & 7 & 7 \\
$w_T$  & Sliding-window length for temporal path  & 31 & 31 & 31 \\
\midrule
$-$  & Number of parameters [M]  & 6.0 & 16.9 & 25.4 \\
\bottomrule
\end{tabular}
\label{tab1}
\vspace{-3mm}
\end{table}
\vspace{-3mm}
\subsubsection{The Convolution Modules}
Local attention captures nearby context but needs many layers for broader integration. To improve global feature mixing efficiently, we add Conv-SwiGLU blocks (Fig. \ref{fig2}(b)), combining attention’s adaptive weighting with convolutional feature extraction. Each block applies RMSGroupNorm, two parallel 1D convolutions, a Swish-gated element-wise modulation, and a final 1D deconvolution to restore embedding size. Following \cite{Kohei2024tflocoformer}, we place Conv-SwiGLU before and after the local and global attention layers.

\section{Experiments}
\label{sec:prior}
\subsection{Dataset \& Experimental Setup}
We validated our proposed TF-MossFormer on the widely-used WSJ0-2mix dataset, which benchmarks monaural speech separation. The dataset is comprised of three distinct sets: a training set of 20,000 (~30 hours), a validation set of 5,000 (~10 hours), and a test set of 3,000 (~5 hours) two-speaker mixtures. The clean utterances are sourced from the WSJ0 corpus. In each mixture, the two utterances are randomly selected from WSJ0 and fully overlapped, with their relative energy levels randomly chosen from the signal-to-noise ratios (SNR) between -5 dB and 5 dB. The test set is generated in the same way. Notably, speakers in the test set are distinct from those in the training and validation sets, ensuring a non-blind evaluation.  All audio clips are sampled at 8 kHz.

All experiments were conducted using the ESPnet pipeline with SI-SDR loss~\cite{li2021espnet}. 
Table~\ref{tab1} summarizes the key hyperparameters and default configurations for our three model sizes.
For data preparation, we used a 16~ms STFT window with an 8~ms hop size and segmented all input audio into 4~s clips. Each mixture was normalized by its standard deviation, and the batch size was set to 4. For optimization, we employed the AdamW optimizer~\cite{Loshchilov2018adam} with a weight decay of $1 \times 10^{-2}$.  The learning rate was linearly warmed up to $1 \times 10^{-3}$ during the first 4{,}000 steps, 
followed by a scheduler that halved the learning rate if the validation loss failed to improve for three consecutive epochs. 
Training was conducted for up to 150 epochs, with early stopping applied after 10 stagnant epochs. 
To ensure stability, the $L_2$ norm of the gradient was clipped to 5.

\subsection{Ablation Study}
\label{sec:prior}
\begin{table}
\centering
\footnotesize
\caption{Ablation study of TF-MossFormer performance with different sliding-window lengths.}
\setlength{\tabcolsep}{0.7pt}
\begin{tabular}{l cccc}
\toprule
Model (S)    & Sliding-Window($w_T, w_F$) & SI-SDRi & SDRi &MACS [G/s] \\
\midrule
TF-MossFormer  & (31, 7) & \textbf{22.61} & \textbf{22.77} &99.5 \\
TF-MossFormer  & (15, 7) & 22.51 & 22.68 &99.4 \\
TF-MossFormer  & (65, 7) & 22.56 & 22.73 &99.7 \\
TF-MossFormer  & (31, 3) & 22.53 & 22.69 &99.5 \\
TF-MossFormer  & (31, 15) & 22.47 & 22.66 &99.5 \\
TF-MossFormer  & Full Length & 22.38 & 22.58 &100.3 \\
\bottomrule
\end{tabular}
\label{tab2}
\vspace{-3mm}
\end{table}

\label{sec:prior}
\begin{table}
\centering
\footnotesize
\caption{Ablation study of TF-MossFormer performance with different configurations of local-global attention.}
\setlength{\tabcolsep}{0.7pt}
\begin{tabular}{l cccc}
\toprule
Model (S)    & Sliding-Window(T-F) & SI-SDRi & SDRi &MACS [G/s] \\
\midrule
TF-MossFormer(V1)  & (31, 7) & \textbf{22.61} & \textbf{22.77} &99.5 \\
TF-MossFormer(V2)  & (31, 7) & 22.45 & 22.60 &99.5 \\
TF-MossFormer(V3)  & (31, 7) & 22.49 & 22.63 &99.5 \\
TF-MossFormer(V4)  & (31, 7) & 22.51 & 22.62 &99.3 \\
\bottomrule
\end{tabular}
\label{tab3}
\vspace{-3mm}
\end{table}
As shown in Table \ref{tab2}, we evaluated several sliding-window configurations for TF-MossFormer (S) on WSJ0-2Mix, keeping the window length identical across all attention heads for ease of tuning. The setting of $w_T=31$ and $w_F=7$ achieves the best result, with 22.61 dB SI-SDRi (22.77 dB SDRi) while maintaining manageable computation at 99.5 G MACS/s, slightly outperforming both the narrower 15-frame and the wider 65-frame time windows. A frequency span of 7 bands is also preferable to smaller (3 bands) or larger (15 bands) choices. Using full-length windows yields the weakest performance (22.38 dB). These results indicate that TF-MossFormer favors moderate time–frequency window sizes that balance local detail and efficiency. Table \ref{tab3} evaluates the local–global attention layouts and convolution gating for TF-MossFormer (S) on WSJ0-2Mix. All models used the (31,7) sliding window. The cascade ordering—local attention followed by global attention (V1)—achieves the best results (22.61 dB SI‑SDRi, 22.77 dB SDRi). Reversing the order (V2) reduces SI‑SDRi to 22.45 dB, while the parallel local–global design (V3) performs slightly better than V2 (22.49 dB). Removing convolutional gating from V1 (V4) also degrades performance to 22.51 dB. These findings suggest that applying local attention before global aggregation helps preserve fine-grained spectral continuity, and that convolutional gating yields measurable gains in feature selection.
\subsection{Experimental Results}
\label{sec:prior}
Our experimental results on WSJ0-2Mix, presented in Tables 4 and 5, comprehensively evaluate TF-MossFormer (configuration V1) across three model scales. The small-scale variant TF-MossFormer(S), with just 6.0M parameters and 99.5 G MACS/s computational cost, achieves 22.61 dB SI-SDRi (22.77 dB SDRi), demonstrating superior separation quality compared to other compact models. It notably outperforms both the standard TF-Locoformer(S) and our reproduced TF-Locoformer(S*) with increased dimensionality ($D=$112), which achieves 22.2 dB SI-SDRi at higher computational cost. The comparison with SPMamba is particularly revealing - TF-MossFormer(S) delivers better performance (by 0.1 dB SI-SDRi) while requiring less than half the computations, highlighting our architecture's efficiency advantages.
At medium scale, TF-MossFormer(M) (16.9M parameters, 283.4 G MACS/s) reaches 24.0 dB SI-SDRi (24.20 dB SDRi), significantly surpassing both TF-GridNet and TF-Locoformer(M) in performance while maintaining substantially lower computational requirements than TF-GridNet. This demonstrates excellent scaling properties of our architecture. The large-scale TF-MossFormer(L) (25.4M parameters, 425.0 G MACS/s) establishes new state-of-the-art results with 24.4 dB SI-SDRi (24.5 dB SDRi), outperforming all frequency-domain competitors including TF-Locoformer(L) and SepTDA$_2$, as well as challenging time-domain baselines like MossFormer and MossFormer2 with dynamic mixing. These consistent gains across all model sizes, from compact to large configurations, robustly validate the effectiveness of our local-global attention design in achieving superior separation quality while maintaining computational efficiency. The results particularly highlight our architecture's advantage in modeling both fine-grained spectro-temporal details and long-range dependencies simultaneously.
\begin{table}
\centering
\footnotesize
\caption{Performance comparison of TF-MossFormer (S) with state-of-the-art models of similar scale on the WSJ0-2Mix dataset.}
\setlength{\tabcolsep}{1.0pt}
\begin{tabular}{l cccc}
\toprule
Model (S)    & Param.[M] & SI-SDRi & SDRi &MACS [G/s] \\
\midrule
Conv-TasNet\cite{Luo2019}  & 5.6 & 15.3 & 15.6 &10.23 \\
\midrule
A-FRCNN\cite{hu2021speech}      & 6.1 & 18.3 & 18.6 &125.3 \\
DualPathRNN\cite{Luo2020Z}  & 2.6 & 18.8 & 19.0 &85.3 \\
VSUNOS\cite{Nachmani2020Y}       & 7.5 & 20.1 & 20.4 &- \\
DPTNet\cite{Chen2020Q}       & 2.6 & 20.2 & 20.6 &102.5 \\
S4M\cite{chen2023neural}          & 3.6 & 20.5 & 20.7 &38.7 \\
SepMamba+DM\cite{avenstrup2024sepmamba}  & 7.2 & 21.2 & 21.4 &- \\
DPMamba\cite{Jiang2025D}      & 8.1 & 21.4 & 21.6 &- \\
TF-Locoformer(S)\cite{Kohei2024tflocoformer} & 5.0 & 22.0  & 22.1 &85.9 \\
TF-Locoformer(S*)\cite{Kohei2024tflocoformer} & 6.0 & 22.2  & 22.3 &100.3 \\
SPMamba\cite{li2024spmamba}     & 6.1 & 22.5  & 22.7 &238.6 \\ 
\midrule
\textbf{TF-MossFormer(S)} & 6.0 & \textbf{22.6}  & \textbf{22.8} &99.5 \\
\bottomrule
\end{tabular}
\label{tab4}
\vspace{-3mm}
\end{table}

\begin{table}
\centering
\footnotesize
\caption{Performance comparison of TF-MossFormer (M) and (L) with state-of-the-art models of similar scale on the WSJ0-2Mix dataset.}
\setlength{\tabcolsep}{1.0pt}
\begin{tabular}{l cccc}
\toprule
Model (M\&L)    & Param.[M] & SI-SDRi & SDRi &MACS [G/s] \\
\midrule
TF-GridNet\cite{wang2023tfgridnet}  & 14.4 & 23.5 & 23.6 &445.5 \\
TF-Locoformer (M)\cite{Kohei2024tflocoformer} & 15.0 & 23.6 & 23.8 &251.1 \\
\textbf{TF-MossFormer(M)}  & 16.9 & \textbf{24.0} & \textbf{24.2} &283.4 \\
\midrule
Wavesplit\cite{Zeghidour2021D}       & 29.0 & 22.2 & 22.3 &- \\
SepFormer+DM\cite{Subakan2021M}       & 25.7 & 22.3 & 22.4 &257.9 \\
SepMamba+DM\cite{avenstrup2024sepmamba}    & 22.0 & 22.7 & 22.9 &- \\
MossFormer(L)+DM\cite{Zhao2023M}  & 42.1 & 22.8 & - &256.7 \\
DPMamba\cite{Jiang2025D}      & 59.8 & 23.4 & 23.6 &- \\
QPDN\cite{rixen22_interspeech}      & 200 & 23.6 & - &- \\
SepTDA$_2$\cite{Lee2024sepTDA}      & 21.2 & 24.0 & 23.9 &- \\
MossFormer2(L)+DM\cite{Zhao2024M} & 55.7 & 24.1  &- &336.9 \\
TF-Locoformer(L)\cite{Kohei2024tflocoformer} & 22.6 & 24.2  & 24.3 &377.1 \\
\midrule
\textbf{TF-MossFormer(L)} & 25.4 & \textbf{24.4}  & \textbf{24.5} &425.0 \\
\bottomrule
\end{tabular}
\label{tab5}
\vspace{-3mm}
\end{table}

\section{Conclusion}

We presented TF‑MossFormer, a hybrid time–frequency transformer that combines adaptive local attention with global attention for monaural speech separation. Our main contributions are: (1) content‑aware sliding‑window attention for adaptive spectro‑temporal modeling; (2) an optimized local–global attention configuration; and (3) convolutional gating to improve feature selection and information flow. Extensive experiments on WSJ0‑2Mix show that TF‑MossFormer delivers state‑of‑the‑art separation performance across multiple model scales while maintaining favorable parameter and computation efficiency compared to existing time‑domain and frequency‑domain baselines.

\bibliographystyle{IEEEtran}
\bibliography{mybib}

@InProceedings{Luo2018Real,
  author = "Y. Luo and N. Mesgarani",
  title = "Real-time single-channel dereverberation and separation with time-domain audio separation network",
  booktitle = "Proc. Interspeech",
  year = "2018"
}

@Article{Luo2019,
  author = "Y. Luo and N. Mesgarani",
  title = "Conv-{T}as{N}et: Surpassing ideal time–frequency magnitude masking for speech separation",
  journal = "IEEE/ACM transactions on audio, speech, and language processing",
  volume = "27",
  number = "8",
  pages = "1256–1266",
  year = "2019"
}

@InProceedings{Erdogan2015,
  author = "H. Erdogan and J. R. Hershey and S. Watanabe and J. Le Roux",
  title = "Phase sensitive and recognition-boosted speech separation using deep recurrent neural networks",
  booktitle = "Proc. ICASSP",
  pages = "708–712",
  year = "2015"
}

@inproceedings{Luo2018N,
  author = "Y. Luo and N. Mesgarani",
  title = "{T}as{N}et: time-domain audio separation network for real-time, single-channel speech separation",
  booktitle = "Proc. of ICASSP",
  year = "2018"
}

@inproceedings{Luo2020Z,
  author = "Y. Luo and Z. Chen and T. Yoshioka",
  title = "{D}ual-{P}ath {RNN}: {E}fficient long sequence modeling for time-domain single-channel speech separation",
  booktitle = "Proc. of ICASSP",
  page = "46–50",
  year = "2020"
}

@inproceedings{Jiang2025D,
  author = "X. Jiang and C. Han and N. Mesgarani",
  title = "Dual-path {M}amba: Short and Long-term Bidirectional Selective Structured State Space Models for Speech Separation",
  booktitle = "Proc. of ICASSP",
  page = "46–50",
  year = "2025"
}

@inproceedings{Nachmani2020Y,
  author = "E. Nachmani and Y. Adi and L. Wolf",
  title = "{V}oice separation with an unknown number of multiple speakers",
  booktitle = "Proc. of ICML",
  page = "7164–7175",
  year = "2020"
}

@inproceedings{Chen2020Q,
  author = "J. Chen and Q. Mao and D. Liu",
  title = "{D}ual-{P}ath {T}ransformer {N}etwork: {D}irect Context-Aware Modeling for End-to-End Monaural Speech Separation",
  booktitle = "Proc. of Interspeech",
  page = "2642–2646",
  year = "2020"
}

@inproceedings{hu2021speech,
      title={Speech Separation Using an Asynchronous Fully Recurrent Convolutional Neural Network}, 
      author={Xiaolin Hu and Kai Li and Weiyi Zhang and Yi Luo and Jean-Marie Lemercier and Timo Gerkmann},
      year={2021},
      booktitle = "Proc. of NeurIPS",
}

@inproceedings{chen2023neural,
      title={A Neural State-Space Model Approach to Efficient Speech Separation}, 
      author={Chen Chen and Chao-Han Huck Yang and Kai Li and Yuchen Hu and Pin-Jui Ku and Eng Siong Chng},
      year={2023},
      booktitle = "Proc. of Interspeech",
}

@Article{Zeghidour2021D,
  author = "N. Zeghidour and D. Grangier",
  title = "{W}avesplit: {E}nd-to-end speech separation by speaker clustering",
  journal = "IEEE/ACM Trans. Audio. Speech, Lang. Process.",
  volume = "29",
  page = "2840-2849",
  year = "2021"
}

@INPROCEEDINGS{Kohei2024tflocoformer,
  author={Saijo, Kohei and Wichern, Gordon and Germain, François G. and Pan, Zexu and Roux, Jonathan Le},
  booktitle={2024 18th International Workshop on Acoustic Signal Enhancement (IWAENC)}, 
  title={{TF}-{L}ocoformer: Transformer with Local Modeling by Convolution for Speech Separation and Enhancement}, 
  year={2024}
}

@inproceedings{Subakan2021M,
  author = "C. Subakan and M. Ravanelli and S. Cornell and M. Bronzi and J. Zhong",
  title = "{A}ttention Is All You Need In Speech Separation",
  booktitle = "Proc. of ICASSP",
  page = "21-25",
  year = "2021"
}

@inproceedings{Vaswani2017N,
  author = "A. Vaswani and N. Shazeer and N. Parmar and J. Uszkoreit and L. Jones and A. N. Gomez and L. Kaiser and I. Polosukhin",
  title = "{A}ttention is all you need",
  booktitle = "Proc. NIPS",
  year = "2017"
}

@inproceedings{Loshchilov2018adam,
  author = "I. Loshchilov and F. Hutter",
  title = "Decoupled weight decay regularization",
  booktitle = "Proc. ICLR",
  year = "2018"
}

@inproceedings{li2021espnet,
  author = "Li, Chenda and Shi, Jing and Zhang, Wei and Subramanian, Anoop S. and Chang, Xugang and others",
  title = "{ESP}net-{SE}: End-to-end speech enhancement and separation toolkit designed for asr integration",
  booktitle = "Proc. SLT",
  year = "2021"
}

@inproceedings{avenstrup2024sepmamba,
      title={Sep{M}amba: State-space models for speaker separation using Mamba}, 
      author={Thor Højhus Avenstrup and Boldizsár Elek and István László Mádi and András Bence Schin and Morten Mørup and Bjørn Sand Jensen and Kenny Falkær Olsen},
      booktitle = "Proc. of ICASSP",
      year={2024},       
}

@Article{wang2023tfgridnet,
      title={{TF}-{G}rid{N}et: Making Time-Frequency Domain Models Great Again for Monaural Speaker Separation}, 
      author={Zhong-Qiu Wang and Samuele Cornell and Shukjae Choi and Younglo Lee and Byeong-Yeol Kim and Shinji Watanabe},
      year={2023},
      journal = "arXiv:2209.03952",
}

@inproceedings{Yang2022L,
  author = "L. Yang and W. Liu and W. Wang",
  title = "{TFPSN}et: Time-Frequency Domain Path Scanning Network for Speech Separation",
  booktitle = "Proc. of ICASSP",
  page = "6842–6846",
  year = "2022"
}

@inproceedings{Lee2024sepTDA,
  author = "Y. Lee and S. Choi and B.-Y. Kim and Z.-Q. Wang and S. Watanabe",
  title = "Boosting Unknown-number Speaker Separation with Transformer Decoder-based Attractor",
  booktitle = "Proc. of ICASSP",
  year = "2024"
}

@Article{li2024spmamba,
      title={{SPM}amba: State-space model is all you need in speech separation}, 
      author={Kai Li and Guo Chen and Runxuan Yang and Xiaolin Hu},
      year={2024},
      journal = "arXiv:2404.02063",
      eprint={2404.02063},
}

@INPROCEEDINGS{Zhao2023M,
  author={Zhao, Shengkui and Ma, Bin},
  booktitle={Proc. of ICASSP}, 
  title={Moss{F}ormer: Pushing the Performance Limit of Monaural Speech Separation Using Gated Single-Head Transformer with Convolution-Augmented Joint Self-Attentions}, 
  year={2023},
}

@INPROCEEDINGS{chen2021continue,
  author={Chen, Sanyuan and Wu, Yu and Chen, Zhuo and Wu, Jian and Li, Jinyu and Yoshioka, Takuya and Wang, Chengyi and Liu, Shujie and Zhou, Ming},
  booktitle={Proc. of ICASSP}, 
  title={Continuous Speech Separation with Conformer}, 
  year={2021},
  volume={},
  number={},
  pages={5749-5753},
}

@Article{zhu2023biformer,
  author  = {Lei Zhu and Xinjiang Wang and Zhanghan Ke and Wayne Zhang and Rynson Lau},
  title   = {BiFormer: Vision Transformer with Bi-Level Routing Attention},
  journal = {Proceedings of the IEEE/CVF Conference on Computer Vision and Pattern Recognition (CVPR)},
  year    = {2023},
}

@Article{ebert2023light,
      title={Light-Weight Vision Transformer with Parallel Local and Global Self-Attention}, 
      author={Nikolas Ebert and Laurenz Reichardt and Didier Stricker and Oliver Wasenmüller},
      year={2023},
      journal={arXiv:2307.09120}, 
}

@article{zaheer2020bigbird,
  title={Big bird: Transformers for longer sequences},
  author={Zaheer, Manzil and Guruganesh, Guru and Dubey, Kumar Avinava and Ainslie, Joshua and Alberti, Chris and Ontanon, Santiago and Pham, Philip and Ravula, Anirudh and Wang, Qifan and Yang, Li and others},
  journal={Advances in Neural Information Processing Systems},
  volume={33},
  year={2020}
}

@article{Beltagy2020Longformer,
  title={Longformer: The Long-Document Transformer},
  author={Iz Beltagy and Matthew E. Peters and Arman Cohan},
  journal={arXiv:2004.05150},
  year={2020},
}

@INPROCEEDINGS{Zhao2024M,
  author={S. Zhao and Y. Ma and C. Ni and C. Zhang and H. Wang and T. H. Nguyen and K. Zhou and J. Yip and D. Ng and B. Ma},
  booktitle={Proc. of ICASSP}, 
  title={Moss{F}ormer2: Combining Transformer and RNN-Free Recurrent Network for Enhanced Time-Domain Monaural Speech Separation}, 
  year={2024},
}

@inproceedings{rixen22_interspeech,
  author={Joel Rixen and Matthias Renz},
  title={{{QDPN} - Quasi-dual-path Network for single-channel Speech Separation}},
  year=2022,
  booktitle={Proc. Interspeech 2022},
  pages={5353--5357},
  doi={10.21437/Interspeech.2022-700}
}

\end{document}